\pdfoutput=1

\documentclass[preprint2]{aastex}

   \usepackage{natbib}

\shorttitle{The Arp~220 merger on kpc scales}
\shortauthors{K\"onig et al.}

\begin{document}


\title{The Arp~220 merger on kpc scales}


\author{S. K\"onig}
\affil{Dark Cosmology Centre, Juliane Maries Vej 30, DK-2100 Copenhagen, Denmark}
\affil{I. Physikalisches Institut, Universit\"at zu K\"oln, Z\"ulpicher Stra\ss e 77, D-50937 K\"oln, Germany}
\email{skoenig@dark-cosmology.dk}

\author{M. Garc\'{i}a-Mar\'{i}n}
\affil{I. Physikalisches Institut, Universit\"at zu K\"oln, Z\"ulpicher Stra\ss e 77, D-50937 K\"oln, Germany}

\author{A. Eckart}
\affil{I. Physikalisches Institut, Universit\"at zu K\"oln, Z\"ulpicher Stra\ss e 77, D-50937 K\"oln, Germany}
\affil{Max-Planck-Institut f\"ur Radioastronomie, Auf dem H\"ugel 69, D-53121 Bonn, Germany}

\author{D. Downes}
\affil{Institut de Radioastronomie Millim\'{e}trique, Domaine Universitaire, F-38406 St. Martin d'H\'{e}res, France}

\and

\author{J. Scharw\"achter}
\affil{Research School of Astronomy and Astrophysics, The Australian National University, Cotter Road, Weston Creek, ACT 2611, Australia}




\begin{abstract}
For the first time we study the Eastern nucleus in greater detail and search for the more extended emission in the molecular gas in
different CO line transitions of the famous ULIRG \object{Arp~220}. Furthermore we present a model of the merger in \object{Arp~220} on large scales with the help of the CO data and an optical and near-infrared composite HST image of the prototypical ULIRG. Using the Plateau de Bure Interferometer (PdBI) we obtained CO\,(2$-$1) and (1$-$0) data at wavelengths of 1 and 3~mm in 1994, 1996, 1997 and 2006 at different beam sizes and spatial resolutions. The simulations of the merger in \object{Arp~220} were performed with the Identikit modeling tool. The model parameters that describe the galaxy merger best give a mass ratio of 1:2 and result in a merger of 
$\sim$\,6$\times$10$^{\rm 8}$~yrs. The low resolution CO\,(1$-$0) PdBI observations suggest that there are indications for emission $\sim$\,10\arcsec towards the south, as well as to the north and to the west of the two nuclei.
\end{abstract}


\keywords{Galaxies: interactions --
                Galaxies: kinematics and dynamics --
                Galaxies: nuclei --
                Radio lines: galaxies}



\section{Introduction}

With an infrared luminosity of L\,=\,1.4\,$\times$\,10$^{\rm 12}$~L$_{\odot}$ at a distance of D$_{\rm L}$\,=\,78~Mpc ($\Omega$$_{\Lambda}$\,=\,0.7, $\Omega$$_{\rm M}$\,=\,0.3, and H$_{\rm 0}$\,=\,70~km\,s$^{\rm -1}$\,M\,pc$^{\rm -1}$), \object{Arp~220} belongs to a class of galaxies known as Ultraluminous IR galaxies (ULIRGs). These types of objects are characterized by having an IR luminosity of 10$^{\rm 12}$\,L$_{\odot}$\,$\leq$\,L$_{\rm bol}$\,$\sim$\,L$_{\rm IR}$\,[8\,-\,1000\,$\mu$m]\,$\leq$\,10$^{\rm 13}$\,L$_{\odot}$. It is not entirely clear yet what exactly the source powering those high luminosities is. Different scenarios have been considered: The nuclear power source may be either a strong nuclear starburst \citep{jos99}, a heavily obscured active galactic nucleus \citep[AGN,][]{san99} or a mixture of the two. Independent of what the true energy source is, a large molecular gas concentration in the central kiloparsecs is needed in order for the energy output to be sustained \citep[see e.g.][]{san88}.\\
\indent
Since ULIRGs as a population show a mixture of starburst and AGN properties, they may trace different phases in a 
ULIRG-to-QSO evolutionary sequence. Various scenarios linking the evolution of ULIRGs to QSOs have been suggested 
\citep[e.g.][]{san88,bek06}. These scenarios allow for a weak, but growing, AGN \citep{tani99} at the center of \object{Arp~220}. 
Theories competing for an explanation of the observed high IR luminosity \citep{sak08} and the rather high star formation rate of 
340~M$_{\odot}$yr$^{\rm-1}$ \citep{baan07} claim, e.g. a pure nuclear starburst contribution, contributions from hot cores or a 
mixture of AGN and starburst. However the detection of Fe K$\alpha$ emission at 6.3~keV at slightly more than 3$\sigma$ significance 
may also indicate the presence of a deeply dust embedded AGN. Observations of the [O{\scriptsize III}]5007 line which is a good AGN 
tracer but also sensitive to star formation, yielded no detection and therefore support the extremely high extinction found for 
\object{Arp~220} \citep[see e.g.][]{coli04}. Optical observations reveal that the galaxy is clearly crossed by a dust lane spreading 
from NE to SW and show the presence of extended tidal tails \citep{arp66} indicating that the galaxy is undergoing a merging event.\\
\indent
Following it's position in the proposed evolutionary ULIRG-to-QSO scenarios, \object{Arp~220} can provide the link between ULIRGs
and elliptical galaxies. It also can be considered as a benchmark for understanding the complex properties of ULIRGs and the objects
representing transition stages to QSOs in detail. \object{Arp~220} may also turn out to be useful to explain the rotation-like
structures detected in high-redshift galaxies \citep[see][]{foer06}. Studying \object{Arp~220} and other ULIRGs in the local universe
in high resolution configurations may allow us to draw conclusions for radio detected galaxies at high redshifts which we cannot
study in this great detail, yet.\\
\indent
High resolution radio and near-infrared imaging revealed a complex nuclear structure \citep[see e.g.,][]{dow07}: The core of
\object{Arp~220} appears to have (at least) two different nuclei \citep[or nuclear regions, e.g.][]{nor88}. They are identified as
Western and Eastern nuclei, separated by about 0.4~kpc. CO\,(1\,-\,0) observations revealed an underlying rotating kpc\,-\,sized molecular gas disk \citep{sco97}.\\
\indent
The brightest source in the complex structure of \object{Arp~220} in the near-IR \citep{sco98} is the western nucleus, 
\object{Arp~220-West}. The high resolution CO\,(2$-$1) map \citep{sco98} and the 1.3~mm continuum maps 
\citep{dow07} show peaks at the same location. The 2.2~$\mu$m/1.1~$\mu$m color map of \citet{sco98} however reveals that the peak of the dust distribution is not associated with this nucleus \citep{dow07}. \citeauthor{dow07} also modeled the CO emission by a compact, hotter (170~K) dust core surrounded by a cooler (50~K) molecular gas ring or disk. They inferred a black body luminosity of the compact dust source 
($\sim$\,10$^{\rm 12}$~$L_{\sun}$), which implies that the energetic process powering \object{Arp~220-West} cannot be of starbursting nature.\\
\indent
The second brightest source in the complex structure of \object{Arp~220} in the near-IR \citep{sco98} and CO emission maps is the eastern nucleus, \object{Arp~220-East}. This nucleus is divided into two components: North-East (CO-NE) and South-East \citep[CO-SE][]{dow07}. \citet{sak09} found a deep absorption in the CO-NE/SE region comparable to the absorption in the CO emission of \object{Arp~220-West} \citep{dow07}. Furthermore, the continuum peaks at the position between CO-NE and SE at the position of the CO-East peak in the velocity dispersion distribution \citep{dow07}, thus implying the region inbetween CO-NE/SE is associated with a deeply embedded and/or 
highly dust obscured nucleus of \object{Arp~220-East}.\\
\indent
The velocity fields obtained from the observational data of the nuclear region show rotation-like patterns at different scales (up
to 1~kpc). By contrast, observations of the outer regions (at scales of $\sim$\,5~kpc) at optical wavelengths present velocity patterns which are disordered and therefore apparently dominated by the merger \citep{coli04}. Influences in the inner regions are shown on the 2~kpc scale where the outflows coming from the dust-enshrouded nucleus affect the ionized gas \citep{arr01}. Velocity field information determined from observations of the CO\,(1$-$0) emission line shows a rotational motion that corresponds to the overall rotation of the underlying molecular disk \citep{sco97}. The velocity field for CO\,(2$-$1) however, looks different: Since the CO\,(2$-$1) line transition probes higher gas densities, one can look deeper into the gaseous environment with this higher molecular transition. Therefore, a more complex rotational pattern emerges: we find indications for rotation in CO-West, CO-NE and SE, with the center located in-between the two intensity peaks \citep[see e.g.,][]{gen01,mun01,rov03,rod05}. Taking the findings for higher resolution CO\,(2$-$1) data into account, one could also understand the larger scale CO\,(1$-$0) velocity field as a combination of the two rotation patterns of the two respective nuclei, but smoothed to a lower resolution. Several models to explain the complex dynamics and kinematics in the molecular gas of \object{Arp~220} have been presented in the literature e.g., the model of one underlying molecular disk \citep{sco97} with additional gaseous disks in the eastern and western nuclei \citep{sak99}. 
\citet{eck01} tried to explain the flux concentrations in \object{Arp~220-West} and East by crowding of inclined circular orbits of a warped disk whereas \citet{dow07} modeled their high-resolution CO\,(2$-$1) observations for \object{Arp~220-West} for the presence of a black hole. They found indications that the western nucleus, in fact, contains a black hole with an accretion disk.\\
\indent
In Sect.\,2 we describe the observations and data reduction, Sect.\,3 reports on the results of the data analysis in 
\object{Arp~220-East}. The results of the analysis of the large scale structure follow in Sects.\,4.
Sect.\,5 concentrates on the dynamical modeling of the merger in \object{Arp~220}.\\
\indent
Unless otherwise stated, $\Omega$$_{\Lambda}$\,=\,0.7, $\Omega$$_{\rm M}$\,=\,0.3, and H$_{\rm 0}$\,=\,70~km\,s$^{\rm -1}$\,M\,pc$^{\rm -1}$ are assumed throughout the paper.

%

\section{Observations and data reduction}


In this paper we work with HST NIR and PdBI\footnote{Based on observations carried out with the IRAM Plateau de Bure Interferometer. 
IRAM is supported by INSU/CNRS (France), MPG (Germany) and IGN (Spain).} CO 1-mm and 3-mm observational data. The line and 
continuum observations of the CO 1~mm and 3~mm emission lines on different spatial scales and resolutions were carried out with the 
IRAM Plateau de Bure interferometer (PdBI) in different configurations during the years 1994, 1996, 1997 and 2006, respectively. The data 
were reduced and analyzed with the help of IRAM's GILDAS\footnote{http://www.iram.fr/IRAMFR/GILDAS} software. Depending on the 
observational setups and the data reduction, the synthesized beam sizes range from 0.30\arcsec\,$\times$\,0.30\arcsec\ to 
4.99\arcsec\,$\times$\,3.50\arcsec. The CO\,(1$-$0) high \citep{dow98} and low resolution data were taken in winter 1996 with a 
bandwidth of 500~MHz and a channel separation of 2.5~MHz, resulting in a velocity range of $\sim$\,1\,300~km\,s$^{\rm -1}$ and a 
velocity channel width of 7~km\,s$^{\rm -1}$. Using natural (high resolution) and uniform weighting (low resolution) in the mapping 
procedure gave beam widths of 1.58\arcsec\,$\times$\,1.11\arcsec\ (high resolution) and 4.99\arcsec\,$\times$\,3.5\arcsec\ (low 
resolution). The HST/WFPC2 F814W archive image was reduced 'on the fly' using the best reference files at the moment of retrieval. 
In this article this image is shown only as a reference. Details on the NICMOS data reduction can be found in \citet{sco98}. 

\begin{figure}[t]
\centering
\includegraphics[width=0.45\textwidth]{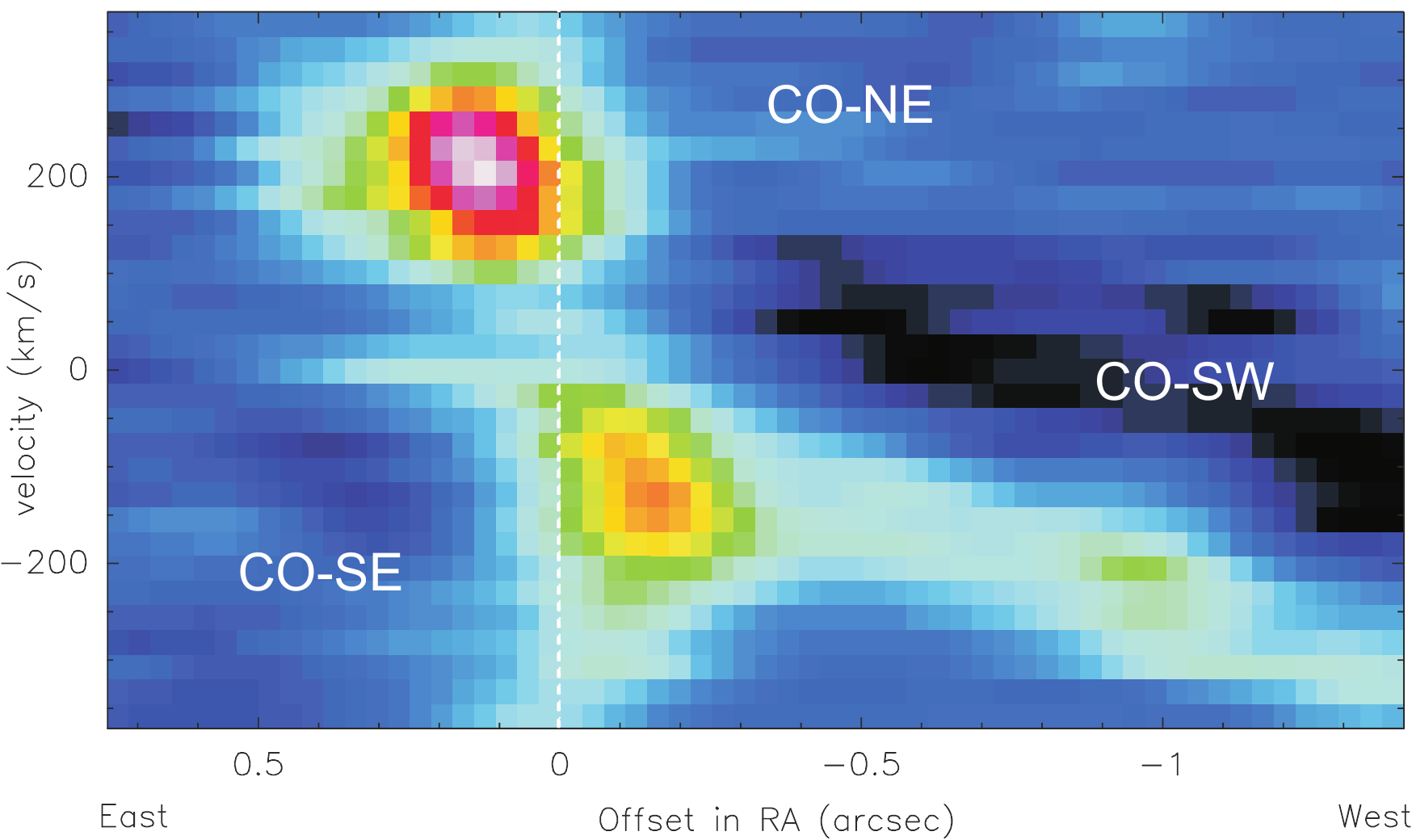}
\caption{\footnotesize East-west (EW) CO\,(2$-$1) position-velocity cut through Arp~220 from CO-NE to CO-SW, with the 1.3~mm
continuum subtracted, considering a position between CO-SE and CO-NE as the rotation center and a radius of 0.3\arcsec.}
\label{fig:CO2-1_pv}
\end{figure} 

\begin{figure*}[ht]
\hspace{-0.8cm}
\centering
\includegraphics[scale=0.6]{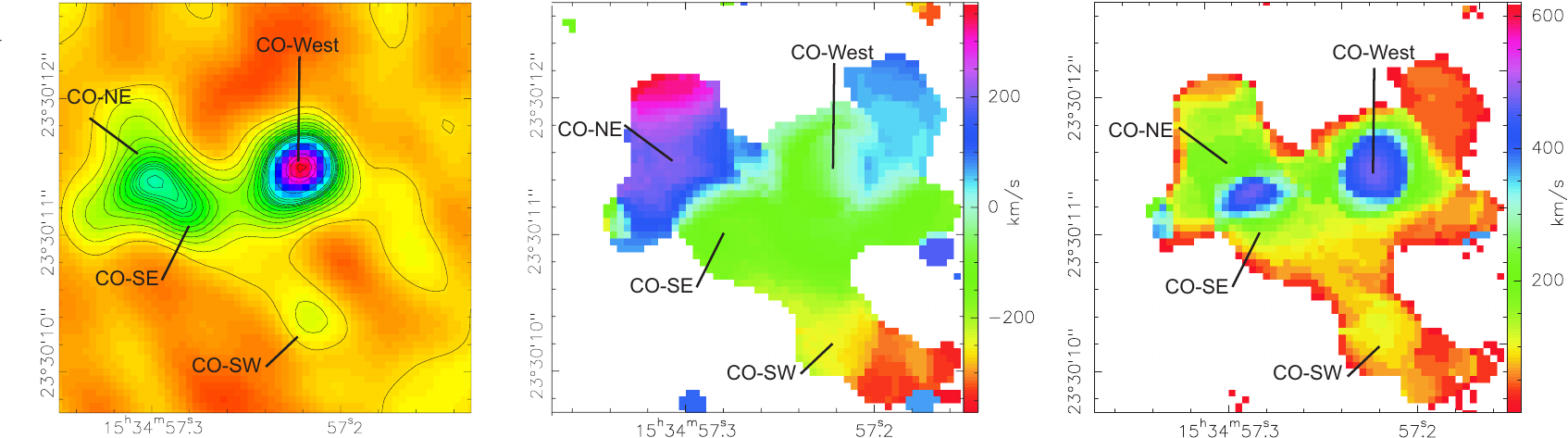}
\caption{\footnotesize CO\,(2$-$1) emission distribution, velocity field and velocity dispersion distribution of Arp~220. 
CO\,(2$-$1) distribution of the central 3\arcsec, with a beam of 0.30\arcsec (\textbf{left}), velocity field (\textbf{middle}) and 
velocity dispersion distribution (\textbf{right}), showing two prominent peaks at the positions of CO-West and CO-East and a possibly fainter third one towards CO-SW.}
\label{fig:CO2-1_velfield}
\end{figure*}

\begin{figure}[t]
\hspace{-0.35cm}
\vspace{-0.1cm}
\centering
\includegraphics[width=0.45\textwidth,angle=-90]{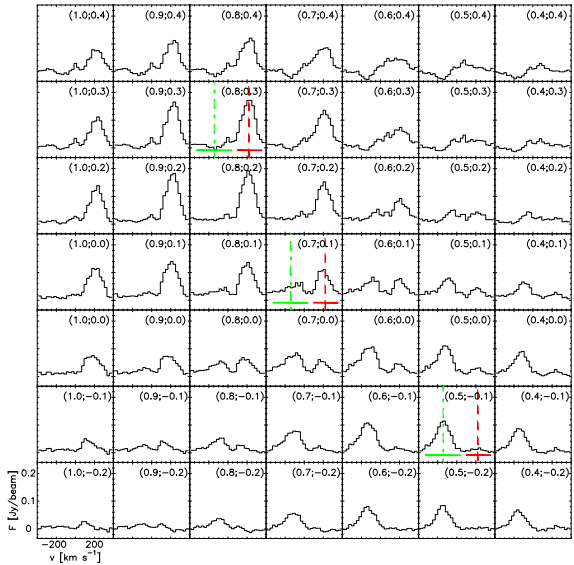}
\caption{\footnotesize CO\,(2$-$1) spectra across Arp 220-East. The position offsets in RA and DEC are marked for each spectrum. The 
spectra are offset by 0.1\arcsec\ from each other. North is up and east to the left. The (0,0) position is centered on the eastern 
velocity dispersion peak (Fig.\,\ref{fig:CO2-1_velfield}). The velocities range from -400 to +400~km\,s$^{\rm -1}$, the intensity range is -0.035 $\leq$ $F$ $\leq$ 0.024 Jy/beam. Note the presence of two velocity components moving from CO-SE
(v$_{\rm 0}$\,$\sim$\,-150~km\,s$^{\rm -1}$, dashed-dotted line) to CO-NE (v$_{\rm 0}$\,$\sim$\,250~km\,s$^{\rm -1}$, dashed line). The bars indicate the widths of the two lines.
}
\label{fig:CO2-1_east_spectra}
\end{figure} 


\section{The kinematics of Arp~220-East}

The eastern nucleus is veiled at optical wavelengths. In the mm wavelengths it can be resolved into a NE and a SE part \citep{dow07}. 
Former publications \citep{dow98,sak99}, before \citet{dow07}, did not resolve the eastern nucleus in two components in CO emission. The velocity gradient which they used to perform their modeling was therefore different from the one shown in Fig.\,\ref{fig:CO2-1_pv} -- in this paper for the first time the CO intensity maps are actually confirmed by the p-v diagram (Fig.\,\ref{fig:CO2-1_pv}). In the near-IR, CO\,(1$-$0) and CO\,(2$-$1) emission maps the eastern nucleus of \object{Arp~220} is the second brightest source after \object{Arp~220-West}. Higher resolution CO observations enabled \citet{sco97} to perform a more detailed comparison between NIR and molecular gas emission: The maximum of the near-IR color map \citep[see Fig.\,2 of][]{sco98} is located at a position between CO-SE and CO-NE and hence indicates that this position is the peak of the dust distribution. Using $H-K$ and a standard extinction law, the $A_{\rm V}$ for the SE and NE components in the NIR are 24 and 18~mag, respectively \citep{rie85}.\\
\indent
The CO\,(2$-$1) spectra for \object{Arp~220-East} (Fig.\,\ref{fig:CO2-1_east_spectra}) show two different velocity components for 
CO-SE and CO-NE. The dominating CO line in CO-SE has a zero-velocity of v$_{\rm 0}$\,$\sim$\,-150~km\,s$^{\rm -1}$, whereas the CO 
emission in CO-NE is centered at v$_{\rm 0}$\,$\sim$\,250~km\,s$^{\rm -1}$. The area between CO-SE and CO-NE marks a transition region in terms of CO line components -- both lines are present but at lower intensities. The p-v diagram (Fig.\,\ref{fig:CO2-1_pv}) 
for \object{Arp~220-East} shows a variation from negative ($v$\,$\sim$\,-200~km\,s$^{\rm -1}$) to positive velocities 
($v$\,$\sim$\,100--150~km\,s$^{\rm -1}$) considering CO-SE as the center of rotation and a radius of about 0.3\arcsec. The CO 
velocity field also shows rotational patterns for the CO-SE/CO-NE region and for CO-SW. The velocity gradients across the regions are 
$\Delta$$v$\,$\sim$\,200~km\,s$^{\rm -1}$ (CO-West), $\Delta$$v$\,$\sim$\,400~km\,s$^{\rm -1}$ (CO-NE \& SE), and 
$\Delta$$v$\,$\sim$\,200~km\,s$^{\rm -1}$ (CO-SW). The velocity dispersion distribution of \object{Arp~220} 
(Fig.\,\ref{fig:CO2-1_velfield} \textit{right}) shows two very distinct peaks, one at the position of CO-West and the other centered 
in the region between CO-SE and CO-NE. The widths of the emission lines rise steeply from $\sim$\,200-250~km\,s$^{\rm -1}$ up to 
$\sim$\,400~km\,s$^{\rm -1}$ in CO-East and even up to $\sim$\,500~km\,s$^{\rm -1}$ in CO-West. Taking the contribution of the underlying disk \citep[visible in CO\,(1$-$0) and (2$-$1) as shown by e.g.][]{sco97,dow98} into account results in velocity dispersions well in agreement with values from e.g. \citet{sco97}.\\
\indent
With the information provided by the CO spectra (Fig.\,\ref{fig:CO2-1_east_spectra}) the emission line width peak being located 
in-between CO-SE and CO-NE (Fig.\,\ref{fig:CO2-1_velfield} \textit{right}) and the structure of the velocity field 
(Fig.\,\ref{fig:CO2-1_velfield} \textit{middle}) it is possible to assert the presence of rotating material around the CO-SE/CO-NE 
region. This seems to point out the CO-SE/NE region as the choice to be identified with a deeply embedded and/or highly dust obscured 
nucleus of \object{Arp~220-East}.\\
\indent
In addition to CO-NE and CO-SE we find that a possible third component, CO-SW \citep[south-west, discovered by][]{dow07}, is
clearly identified in the near-IR \citep{sco98} and in CO\,(2$-$1) (line width enhancement $\sim$\,50~km\,s$^{\rm -1}$ compared to
the surroundings of CO-SW, Fig.\,\ref{fig:CO2-1_velfield} \textit{right}), although it is not detected in the 1.3~mm continuum \citep[see Fig.\,2 of][]{dow07}.\\
\indent
The extinction values derived for CO-SW from the NIR are $A_{\rm V}$\,=\,17~mag. The structure of the velocity dispersion
map (Fig.\,\ref{fig:CO2-1_velfield}) suggests an additional peak towards CO-SW. This enhancement could be produced by a mass concentration, or alternatively by turbulent motions due to the merger process. The velocity field of CO-SW shows 
a rotation pattern centered in this region, which is also reflected by the local gradient in the p-v diagram 
(Figs.\,\ref{fig:CO2-1_pv}, \ref{fig:CO2-1_velfield}). Although the true nature of CO-SW is 
still unclear, it could be an additional very faint nucleus of a possible minor merger component (mergers that involve a gas-rich 
disk galaxy and a bound companion or satellite that usually has $\lesssim$\,10$\%$ of the mass of the gas-rich galaxy), a huge star 
forming region or a remnant of the merging process from one of the colliding galaxies.\\
\indent
In the simulations of the merger in \object{Arp~220} (see Sect.\,\ref{section:identikit}), the CO-SE and NE peaks are handled as
the core of merger component II, the western nucleus is the core of merger component I.
\begin{figure}
\centering
\includegraphics[width=18pc]{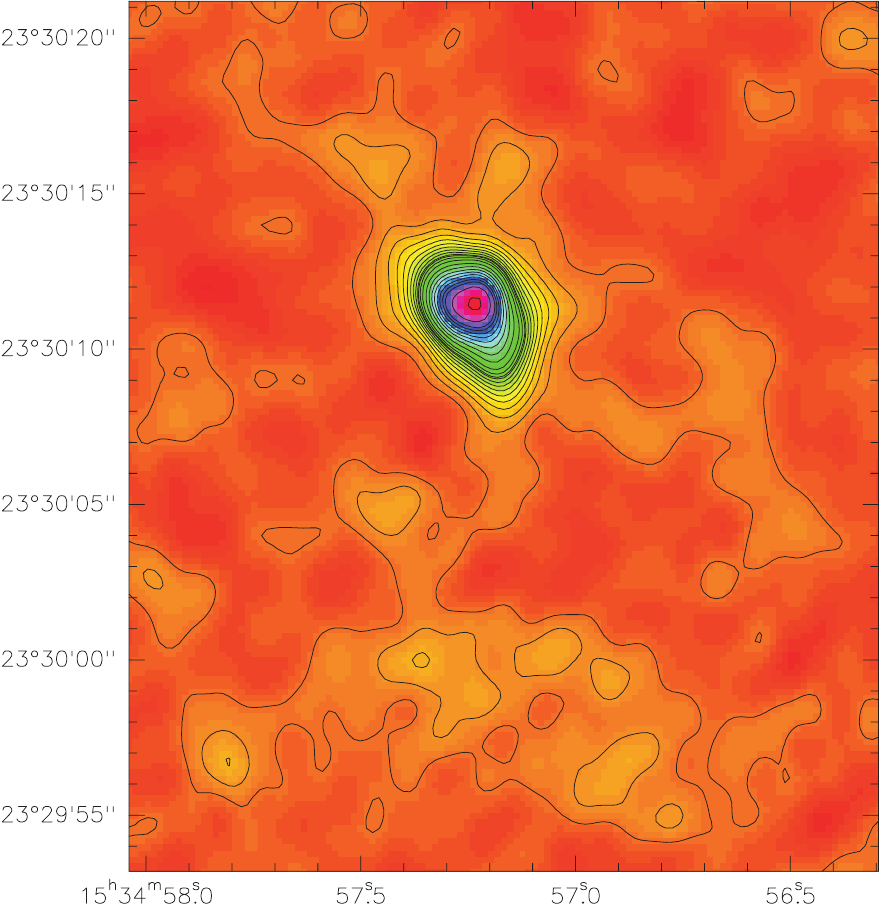}
\caption{\footnotesize Integrated intensity map of the CO\,(1$-$0) emission in Arp~220 integrated over a velocity range of
730~km\,s$^{\rm -1}$ with a beam size of 1.57\arcsec\,$\times$\,1.11\arcsec and with contours starting at the 1$\sigma$ level with
increasing intensity in steps of 2$\sigma$.}
\label{fig:CO1-0_intensity_map}
\end{figure}

\begin{figure}
\centering
\includegraphics[scale=0.4]{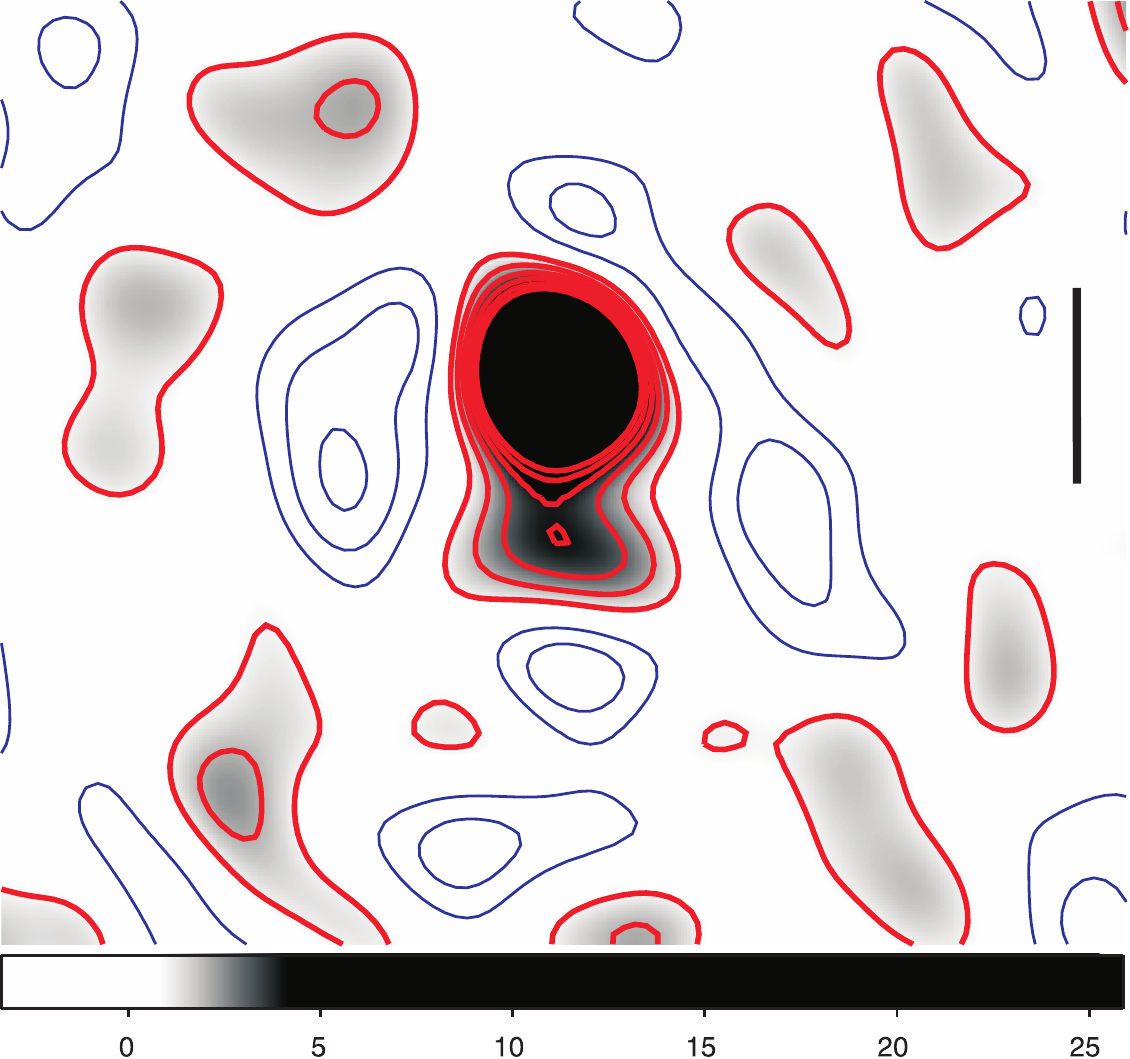}
\caption{\footnotesize CO\,(1$-$0) emission distribution smoothed convolved with a 7.8\arcsec\ beam, centered on the position of Arp~220. Positive and negative contour levels (indicated in red and blue) at steps of 1.25$\sigma$, excluding 0. The grey-scale indicates the intensity. The bar represents a size of roughly 10~kpc.}
\label{fig:largescale_discussion}
\end{figure}

\section{The large scale picture}

The structure of the central region of \object{Arp~220} has been studied extensively over the years. 
These studies \citep[see e.g.,][]{sco97,sak99,dow07} have shown that
the CO and continuum brightness imply a total gas mass ($H_2$+$H$) that approaches 10$^{10}~M_{\sun}$
and a mean nuclear $H_2$ mass density around several 1000 $M_{\sun}~pc^{-3}$.
The total mass of the eastern and western nucleus is still under discussion. It depends
critically on whether the dynamical mass is totally dominated by the molecular gas
or rather the stellar and possibly (at least in the western nucleus) black hole mass contribution.
The best one can say right now is that the masses of the two nuclei (excepting the larger 
scale molecular disk they are located in) may be similar to within a factor of two (or more).
Little effort, however, has been dedicated to study the more external regions although 
there are indications for star formation activity in these regions, too.\\
\indent
The elliptical-like envelope of the underlying molecular disk \citep[see e.g. the CO\,(1$-$0) velocity field map in][]{dow98} may be consistent with a late merger phase within the elliptical-through-merger scenario. The
fact that near-IR light profiles follow an r$^{\rm 1/4}$-law \citep{wri90} corroborates this statement. This law found by 
\citet{deV53} describes the characteristic radial dependence of the surface brightness distribution of elliptical galaxies.\\
\indent
In this paper we study the molecular gas outside the central 3\arcsec, which has not been subject to a detailed analysis in the 
previous literature. In CO\,(1$-$0) we find indications for emission extending about 10\arcsec\ to the south as well as extended 
emission to the north and west of the CO-E and CO-W components (Fig.\,\ref{fig:CO1-0_intensity_map}). Our recent more careful 
re-analysis of the data used in \citet{dow98} makes us confident, that the detection is real. A further comparison with other 
low-resolution CO observations yielded the same result.\\
\indent
The molecular mass for the extended emission towards the south of \object{Arp~220} is determined from:
\begin{equation}
\label{equ:nh2}
M = \langle N \bigg [ \frac{H_{\rm 2}}{X} \bigg ] \rangle \cdotp A \cdotp 2 \cdotp \mu \cdotp m_{\rm H}\ \ M_{\sun}\ \ .
\end{equation}
$M$ is the total mass of molecular hydrogen of the extended emission region, $\langle N(H_{2})\rangle$ is the hydrogen column
density determined from the low-resolution CO\,(1$-$0) data, $A$ is the area projected onto the sky in the distance $D_{\rm L}$
in cm$^{\rm -2}$, $m_{\rm H}$ is the mass of one hydrogen atom (1.67$\times$10$^{-24}$~kg) and $\mu$ is the average molecular weight
per hydrogen atom ($\mu$\,=\,1.36). Applying this formula to the CO data for this region results in a total molecular gas mass (within the 1$\sigma$ level contour line) 
of $\sim$\,1\,$\times$\,10$^{\rm 8}$~$M_{\sun}$.\\
\indent
In Fig.\,\ref{fig:largescale_discussion} we show the line intensity map convolved with a 7.8\arcsec\ circular beam. The southern
component appears as a natural extension of the central source. A few shallow negative regions surround the source plus southern
extension symmetrically at a radius of about 20\arcsec\ followed by a positive noise belt at a radius of 30\arcsec.\\
\indent
The main source plus its southern extension lies at the center of those rings. The symmetry of the first negative ring supports the
significance of the extended positive source structure in the center. At that resolution, at which the source size of the southern
extension is well met, the positive southern extension reaches 5$\sigma$.


\section{Dynamical modeling of the merger} \label{section:identikit}

\object{Arp~220} is the archetypical infrared luminous galaxy merger. It shows tidal tails and a disturbed central region hosting
two prominent nuclear sources. Motivated by the detection of molecular gas in the more extended host-galaxy regions of \object{Arp~220} and the two nuclei, we present a merger model based on the collisionless, non-selfgravitating multiparticle code 
Identikit \citep{bar09}. Before describing the details of the modeling we justify why a simulation with a  collisionless multi-
particle code can be used despite the apparently large mass of dense molecular material at the merger nucleus.

\subsection{Motivation}

While the nuclear region of \object{Arp~220} contains significant amounts of dense molecular material, it encompasses only a few per mill at most of the entire volume over which the merger takes place and has been modeled by us. For the dominant portion of the host we assume as a working hypothesis that across the \object{Arp~220} merger the molecular gas predominantly occurs in clumps with
sizes of about 1~pc \citep[sizes of 10$^{0\pm1}$~pc are quoted by e.g.][]{par11,fon02,hof00,val00}.\\
\indent
With the help of the mass of the gas clumps and the total gas mass of the extended region to the south of \object{Arp~220} the
particle collision rate can be estimated. Using the critical density commonly found in gas clumps
($n_{\rm crit}$\,=\,10$^{\rm 3}$\,...10$^{\rm 5}$~cm$^{\rm -3}$), the clump mass ($M_{\rm clump}$) can be derived by
\begin{equation}
 M_{\rm clump} = n_{\rm crit} V m_{\rm H} ~\big[M_{\sun}\big] .
\end{equation}
$V$ is the volume of one clump ($\sim$\,1~pc$^{\rm 3}$), $m_{\rm H}$ (1.67\,$\times$\,10$^{\rm -27}$~kg) is the mass of one H atom and $M_{\rm clump}$ is given in units of solar masses M$_{\sun}$. 
Since we calculated the gas mass of the southern region of extended
CO emission (M$_{\rm gas}$\,$\sim$\,10$^{\rm 8}$~M$_{\sun}$), the clump density can now be determined, via
\begin{equation}
 M_{\rm clump} \int N dV = M_{\rm gas}
\end{equation}
where $M_{\rm clump}$ and $M_{\rm gas}$ are known, $N$ is the clump density and $dV$ is the volume of the southern region of extended
CO emission. With $N$, the velocity dispersion $\sigma$ ($\sim$\,10~km\,s$^{\rm -1}$) and the cross-section $s$ 
($\sim$\,1~pc$^{\rm 2}$) known, finally the collision rate $R_{\rm coll}$ can be derived by:
\begin{equation}
 R_{\rm coll} = N\,\sigma\,s.
\end{equation}
From this value the time $t_{\rm coll}$ between two collisions is then determined by the reciprocal of the collision rate. This value gives an upper limit for $t_{\rm coll}$ of particles. Values for the time scale range between 3.6\,$\times$\,10$^{\rm 8}$ and 
3.6\,$\times$\,10$^{\rm 10}$~yrs, dependent on the critical density of the gas clumps used in the calculations. These time scales are comparable to the dynamic time scale of the merger itself. The other values for the time between collisions are even larger. Even if agglomeration between the gas clumps should occur, the energy contribution from the merger process, via tidal forces, to the velocity dispersion of the gas is large enough that the clumps in the outer regions do not collide any more. Therefore the gas clumps can be treated, to first order, as stars. Both, stars and gas clumps can then be handled, to first order, as collisionless particles. This fact is implemented in the Identikit code, so that it is justifiable to use this modeling tool for the purposes of the simulation presented. For that particular reason the Identikit model can be used without restriction.\\
\indent
Also the contribution of the atomic gas component can be neglected: Since \object{Arp~220} is already in the final stage of merger the interacting galaxy component particles already collided, in the center of \object{Arp~220}, at least 1--2 times with each other. Therefore the HI is either stripped from the region studied with the merger simulations \citep[Fig.\,3 of][]{hib00} or is already transformed into molecular gas (H$_{\rm 2}$).

\begin{figure*}[!ht]
\vspace{-2.0cm}
\begin{center}
\includegraphics[scale=0.5]{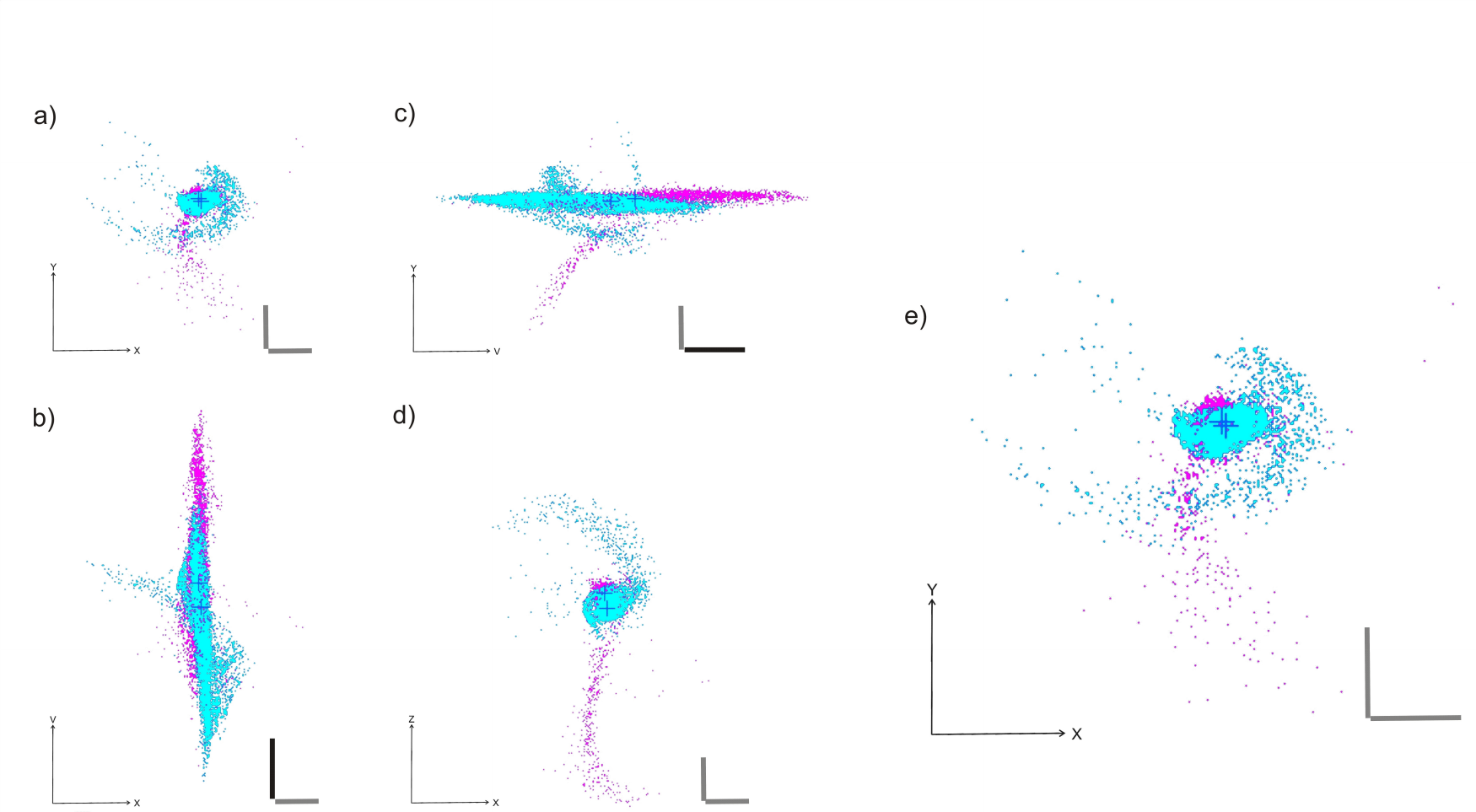}
\caption{\footnotesize Identikit results for the model with the best fit to the available data of Arp~220. a) to d)
show the model projected onto the x-y (a), x-v (b), v-y (c) and x-z (d) planes. In e) the magnified x-y plane is shown. The grey bar
represents a size of roughly 10~kpc and the black bar represents a velocity scale of $\sim$\,500~km\,s$^{\rm -1}$.}
\label{fig:model_end}
\end{center}
\end{figure*}

\subsection{The Identikit model}

The Identikit model \citep{bar09} is a tool to simulate mergers of galactic disks with the help of test particles. Test particles
(representing stars and clumps in this simulations) are used due to their ability to reproduce features such as bridges, tails and
shells, in particular, very well. The simulations described by the Identikit model are based on collisionless N-body simulations
(for the validity of the assumption of collisionless particles see Sect.\,\ref{subsection:identikit_discussion}). The simulations
presented here are based on a model where a bulge, a disk, and a halo component are included.\\
\indent
The initial orientation of the two galaxy disks related to the orbit is described by the inclination \textit{i} and the peri-centric 
argument \textit{$\omega$} of each galaxy disk. Parameters defining the orbit are the eccentricity of the orbit \textit{e}, the 
peri-centric separation \textit{p} and the mass ratio \textit{$\mu$} of the two disks. In the Identikit model the eccentricity is 
fixed at a value of one. The evolution time \textit{t} describes the age of the merger in internal units of 
$-$2\,$\leq$\,\textit{t}\,$\leq$\,8. \textit{t}\,=\,0 represents the point in the merger evolution at which, in the configuration of 
the merging/interacting system, the idealized Keplerian orbit reaches the peri-center.

\subsection{Simulation results}

\begin{figure*}[t]
\vspace{-2.0cm}
\centering
\includegraphics[scale=0.7]{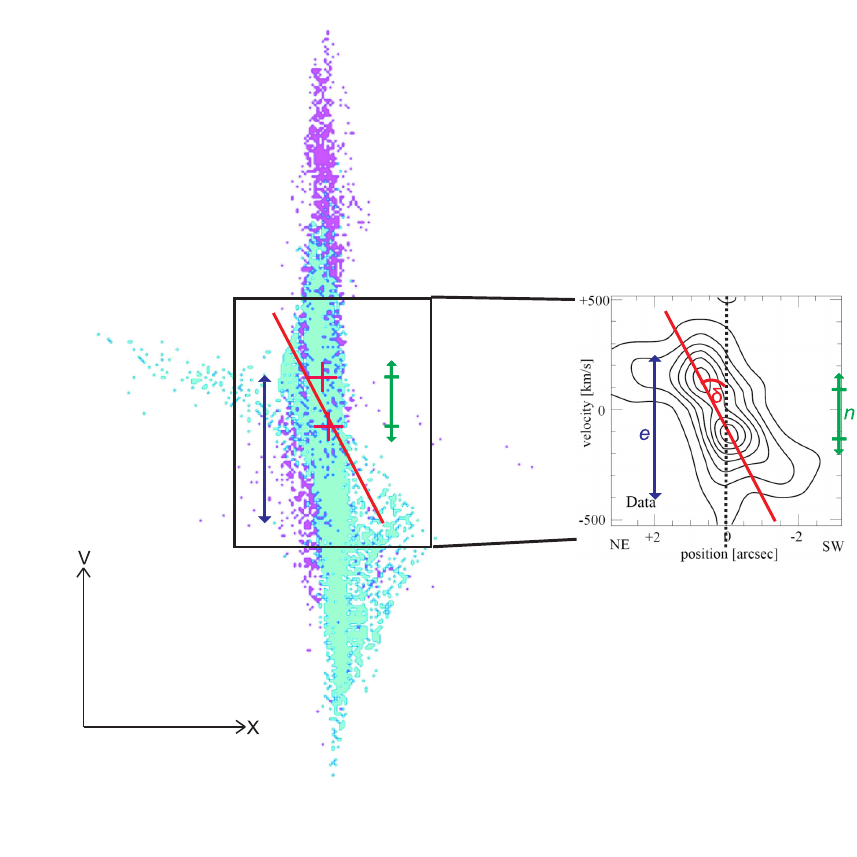}
\caption{\footnotesize Comparison of the position-velocity diagram obtained from the Identikit simulations (\textit{left}) with the p-v diagram from CO\,(2$-$1) observations (\textit{right}). The quality test parameters $n$ (in green), $e$ (in dark blue) and $\delta$ (in red) are indicated in both plots. Note that the CO p-v diagram does not cover the same velocity range as the simulations do, since it was obtained for the center of Arp~220 and does not take the regions furthest outwards into account.}
\label{fig:vergleich_pv}
\end{figure*}

\begin{figure*}
\vspace{-2.0cm}
\begin{center}
\includegraphics[scale=0.5]{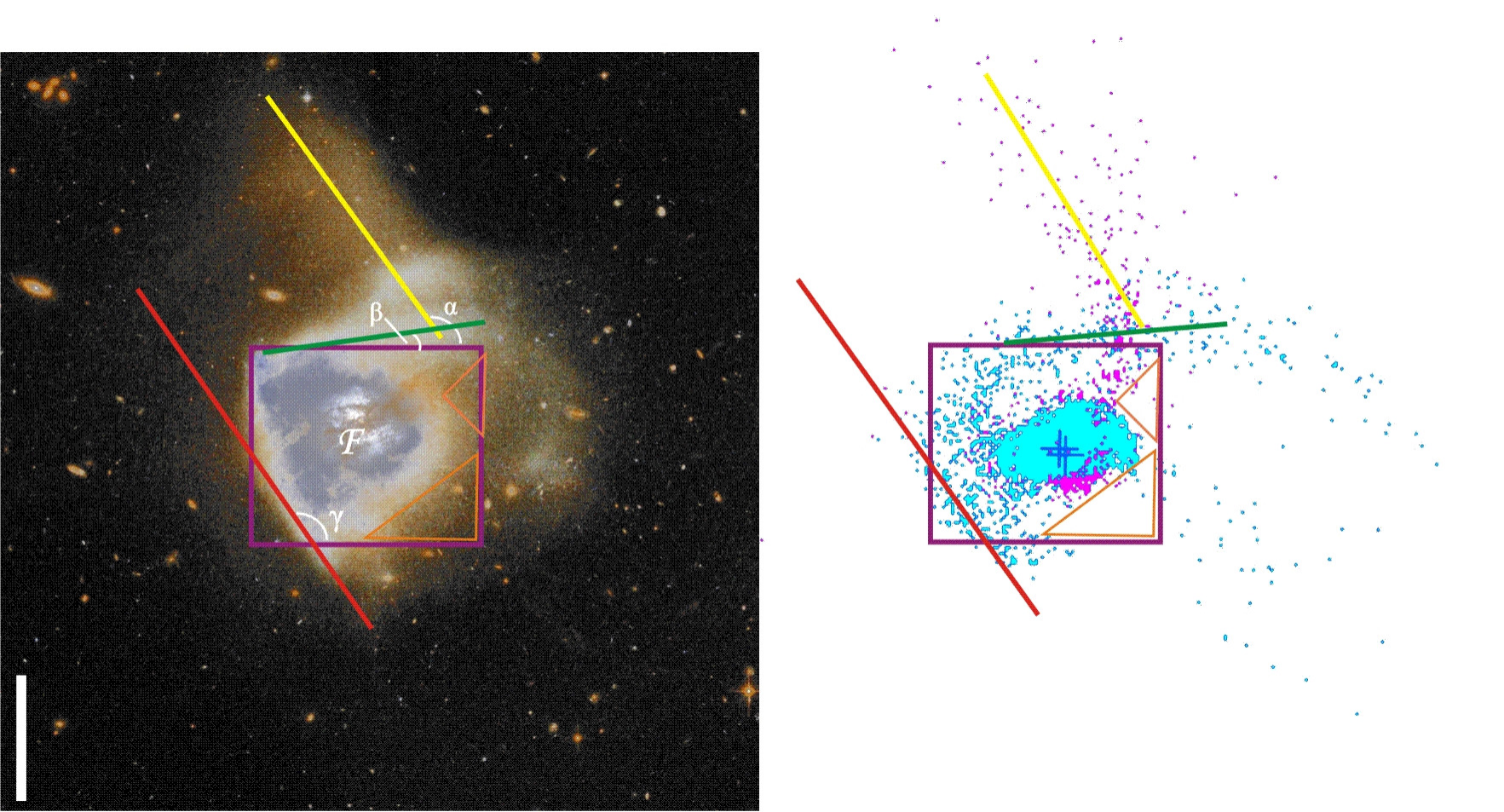}
\caption{\footnotesize The model evaluation criteria are shown on top of the HST image (\textit{left}) and the best model fit
(\textit{right}) for a comparison of the observed data to the model fit and an assessment of the quality of this model. $\alpha$,
$\beta$ and $\gamma$ represent the position angles of the two tidal tails and the sharp edge of Arp~220 to the east,
respectively. $F$ is the surface filling factor, which is a measure for the size and compactness of the galaxy. The bar represents a size of roughly 10~kpc.}
\label{fig:criteria}
\end{center}
\end{figure*}

Optical HST archive composite image\footnote{from \textit{http://imgsrc.hubblesite.org/hu/db/images/hs-2008-16-aq-full\_jpg.jpg (Credit: NASA, ESA, the Hubble Heritage (STScI/AURA)-ESA/Hubble collaboration, and A. Evans (University of Virginia,
Charlottesville/NRAO/Stony Brook University))}} and mm-CO \citep[obtained by][]{eck01,sco97} data are used to fit the Identikit
model simulations. The HST images were used to identify the 'overall shape' (x-y plane) and the CO\,(2$-$1) data were used to determine
a measure on the position-velocity diagrams (x-v plane).\\
%
\begin{figure*}[ht]
  \begin{minipage}[hbt]{0.5\textwidth}
  \centering
    \includegraphics[width=0.9\textwidth]{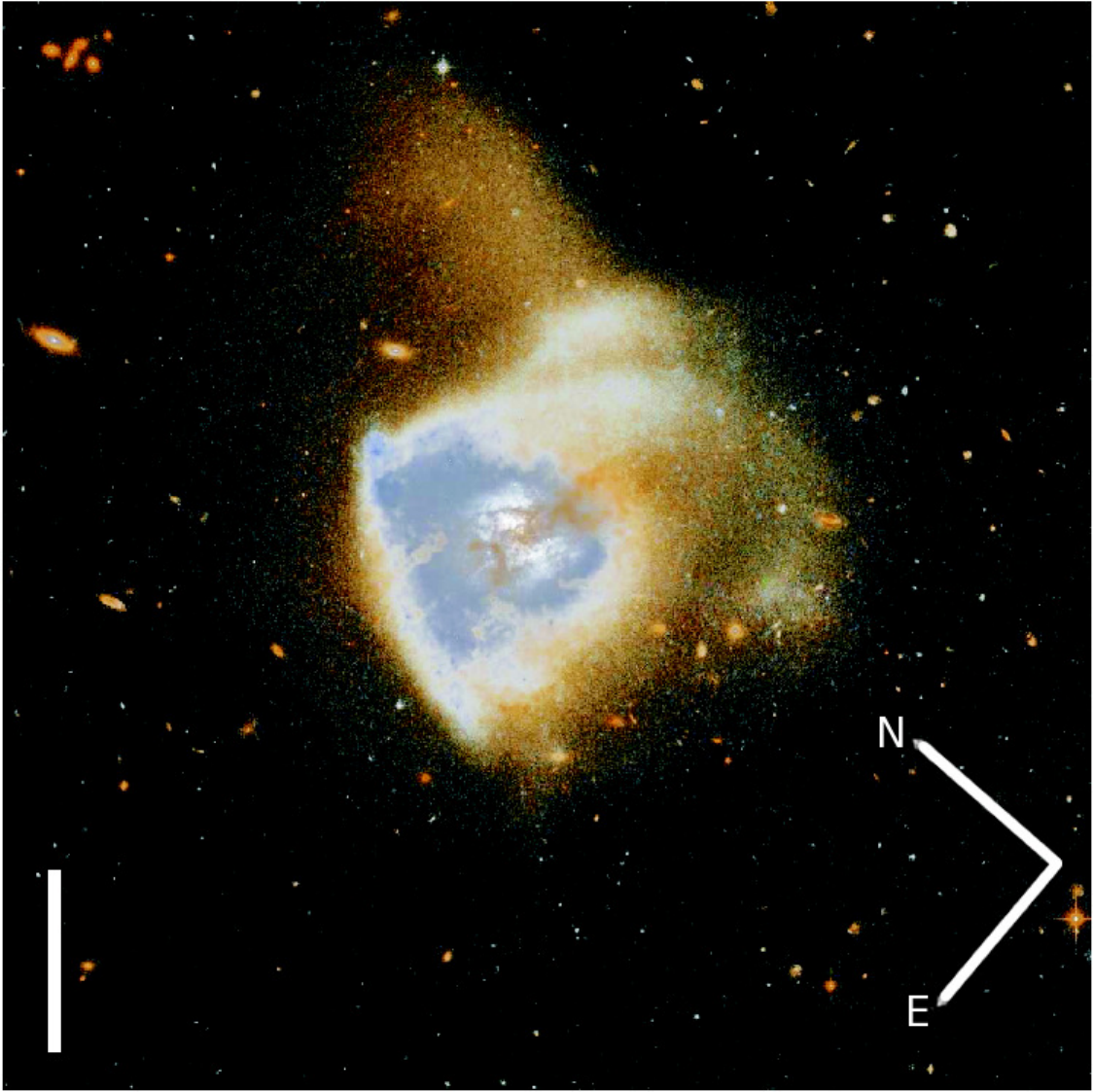}
  \end{minipage}
  \begin{minipage}[hbt]{0.5\textwidth}
  \centering
    \includegraphics[width=0.9\textwidth]{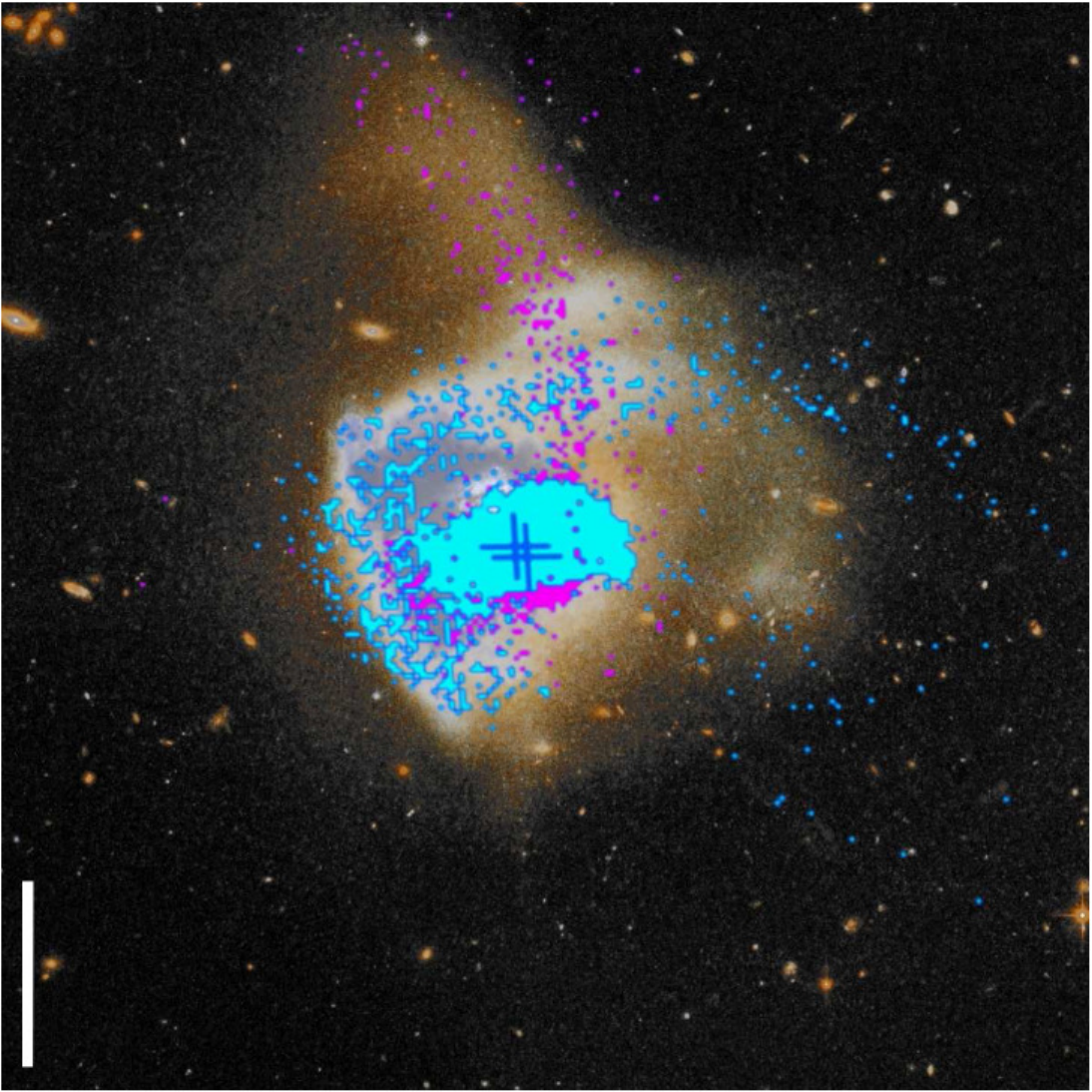}
  \end{minipage}
  \caption{\footnotesize Arp~220 HST image overlay with the best model. \textbf{Left:} HST ACS/WFC F435W and F814W composite image of Arp~220 with an increased contrast in order to show the tidal tails of the galaxy more clearly. \textbf{Right:} The same, but superimposed with the x-y plane of the best model fit and also with an increased contrast (between model and background image). The white bar represents a size of roughly 10~kpc.}
  \label{fig:overlay}
\end{figure*}
\indent
In order to obtain a grid of possible models and to obtain the 'best model fit' from that grid, many different simulations were run systematically. To get a first estimate on the model parameters for \object{Arp~220}, the model of the \object{Antennae Galaxy} was used as a starting point, especially for the inclinations \textit{i} and pericentric arguments \textit{$\omega$} of the galaxy disks \citep[\object{Antennae}: $i_{\rm 1}$\,=\,$i_{\rm 2}$\,=\,60\degr, $\omega_{\rm 1}$\,=\,$\omega_{\rm 2}$\,=\,-30\degr,][]{toomre72,bar88}. The \object{Antennae Galaxy} has a similar morphology, compared to \object{Arp~220}, and it also shows tidal tails from both merging galaxies, but the viewing angle is different. The eccentricity was fixed internally from the start at a value of $e$\,=\,1. In the Identikit simulation tool, only two values for the mass ratio $\mu$ were available to choose from, 1:1 and 1:2
with the eastern nucleus being the mass component 1.
 Actually, the simulations were performed for both mass ratios. At the start of the simulations all possible values for the pericentric separation $p$, in the combination with the merger time scale $t$, were run. After the first few crudely determined simulations, the parameters, $e$, $\mu$, $p$, $i_{\rm 1}$, $\omega$$_{\rm 1}$, $i_{\rm 2}$, $\omega$$_{\rm 2}$ and $t$, were better constrained to a smaller parameter space, the viewing angles were varied. First only one of the angles was changed, the other two remained fixed, later all were changed in combination. The variation of the viewing angles was performed systematically in steps of 10\degr.\\
\indent
Changing the parameters and the combinations thereof resulted in a set of parameters describing the model that fits the observational data best (Fig.\,\ref{fig:model_end}). This best fitting model was obtained by a visual inspection of the model fits and the observational data and by applying a set of different criteria (see also Fig.\,\ref{fig:criteria}).
This best fitting model has a point in the time evolution that 
results in a good match between the position and velocity of the 2 nuclei 
(Figs.\,\ref{fig:model_end}, \ref{fig:overlay} and \ref{fig:vergleich_pv})
 and represents essential features of the more extended distribution of the molecular gas
(Figs.\,\ref{fig:overlay} and \ref{fig:criteria}).
\indent
\\
The set of Identikit model parameters delivering the best solution fitting of the HST image and the CO position-velocity diagram is comprised of the following parameters: The eccentricity at a value of $e$\,=\,1. For the pericentric separation \textit{p} a value of 0.125 delivered the best results. A mass ratio of the two galaxy disks of 1:2, an unequal merger (see Figs.\,\ref{fig:model_end} and \ref{fig:overlay}) resulted in the best model fit. The lower mass galaxy (see Fig.\,\ref{fig:model_end} in magenta) was initiated with an inclination $i$ of 60\degr\ and a pericentric argument $\omega$ of $-$15\degr, the higher mass galaxy was described initially by an inclination $i$ of $-$60\degr\ and a pericentric argument $\omega$ of 40\degr. The viewing angles of $\Theta_{\rm X}$\,=\,90\degr, $\Theta_{\rm Y}$\,=\,180\degr\ and $\Theta_{\rm Z}$\,=\,40\degr\ gave a good match of the simulation data to the observations. The evolution time scale for the merger resulted in a value of $t$\,=\,3 (in internal units). 1$\sigma$ errors of the Identikit modeling parameters are of the order of 8.3\degr\ for the viewing angles $\Theta_{\rm X}$, $\Theta_{\rm Y}$ and $\Theta_{\rm Z}$. The errors for the internal time scale, the inclinations and pericentric arguments of the disks are $\Delta t$\,$\sim$\,0.1, $\Delta i$\,$\sim$\,10\degr\ and $\Delta \omega$\,$\sim$\,10\degr. A comparison of the obtained model fit of \object{Arp~220} to the study of \citet{gon05}, of merging galaxy systems, results in a translation of the evolution time \textit{t} to a merger age of roughly $\sim$\,6\,$\times$\,10$^{\rm 8}$~yrs. This value is in very good agreement with the age of the merging system determined by \citet{mun01}. They give a value of 7\,$\times$\,10$^{\rm 8}$~yrs.\\
\indent
The comparison of the HST image and the best model fit in Fig.\,\ref{fig:overlay} shows that the model very well represents the
properties of the \object{Arp~220} in the HST observations. The sharp edge towards the north-east is well represented by the model. The tidal tail of the less massive galaxy (in magenta) is in agreement with the extended emission towards the north-west. The more southern tidal tail is, at the basis towards the central region of \object{Arp~220}, in good
agreement with the HST image. Further out this extended emission does not coincide with the model any more. As shown in
Fig.\,\ref{fig:vergleich_pv}, the overall shape of the model matches the CO p-v diagram. Note that a difference is that the CO p-v diagram was taken in the central region of \object{Arp~220}. Hence it does not cover the same velocity range as the simulations do and does not take the regions furthest outwards into account (see Fig.\,\ref{fig:vergleich_pv}). The extensions to the NE and the SW side of the zero-offset axis of the p-v diagram describe the CO p-v diagram very well. The difference in the velocity range originates from the test particles further outward of the central region not covered by the CO data. Also the extended emission found in the low resolution CO observations (Fig.\,\ref{fig:CO1-0_intensity_map}) coincides very well with the simulations: At the location of the extended emission $\sim$\,10\arcsec\ to the south of the central part of \object{Arp~220} a high test particle density is found in the Identikit simulations (Fig.\,\ref{fig:model_end}).
It is also worth noting that in Fig.\,\ref{fig:vergleich_pv}
the density of the simulation dots on the left does neither reflect
the molecular gas mass solely nor does it reflect the emissivity of the molecular gas.
Also the right hand image is spatially convolved with the beam, whereas the left hand
site is not convolved to allow for a clearer view onto the distribution of
simulation dots.
\\
\indent
The quality of the model fit was determined by the following criteria: Size and compactness of the merger galaxy, and by the
location and position angle of the tidal tails. The position angles, $\alpha$ and $\beta$, of the tidal tails to the north-west
(represented by the yellow line in Fig.\,\ref{fig:criteria}) and closer to the central region (represented by the green line in
the same Figure) of \object{Arp~220} have values of $\alpha$\,$\sim$\,(124.6\,$\pm$\,9.6)\degr\ and 
$\beta$\,$\sim$\,(6.9\,$\pm$\,7.6)\degr. The sharp edge of the central region of \object{Arp~220} is located at a position angle 
$\gamma$ of $\sim$\,(125.2\,$\pm$\,4.8)\degr. The surface filling factor, a representation of the size and compactness of the galactic central bulge, was found to be $F$\,$\sim$\,(79\,$\pm$\,6)$\%$. This filling factor $F$ represents the area within the box, shown in Fig.\,\ref{fig:criteria}, occupied by the central region of \object{Arp~220}. $F$ is determined by fraction of the area of the box and the box area, with the triangles subtracted where the particle density is reduced to about zero:
\begin{equation}
F = \frac{box\ area - triangles}{box\ area} \ \ \ \%
\end{equation}
All errors stated for the results are errors at the 1$\sigma$ level. By direct comparison, this 1$\sigma$ error for each criteria
parameter was determined from the HST image. For the error values resulting thereof a $\chi$$^{\rm 2}$ (Eq.\,3.1) of 1.3 was derived by
\begin{equation}
 \chi^{\rm 2} = \sum_{i=1}^{N_{\rm obs}} \frac{\big( f_{\rm obs,i} - f_{\rm model,i} \big)^{\rm 2}}{\sigma_{\rm i}^{\rm 2}}
\end{equation}
as a measure for the deviation of the best model fit from the observed data. $f_{\rm obs,i}$ is the respective criteria parameter
measured in the observational data, $f_{\rm model,i}$ is the parameter determined from the best model fit and $\sigma_{\rm i}$ is the observational error. Assuming $\Delta \chi ^{\rm 2}$\,$\sim$\,1 for the variation of each Identikit parameter separately (one
parameter was variable, the others were kept fixed), the internal errors for the Identikit model parameters were derived (see
before). 10 model parameters in the Identikit tool are faced by 10 criteria parameters obtained from the observations: location
(six parameters) and position angle (three parameters) of the tidal tails and the sharp edge of \object{Arp~220}, and the surface
filling factor (one parameter). With the help of these criteria the quality of the fitted models was assessed. The comparison of
the best model to the HST image in Fig.\,\ref{fig:criteria} shows how well it fits the observational data.\\
\indent
We furthermore check the quality of the identikit model fit versus the CO p-v diagram. We therefore use three parameters to compare the CO p-v diagram to the p-v diagram obtained from the identikit simulations: $n$ (in green) represents the distance between the two nuclei, $e$ (dark blue) is a measure for the distance between the NE and SW extensions, relative to the zero-offset axis and $\delta$ (red) describes the angle between the zero-offset axis and the axis on which the two nuclei are situated. All three parameters are in good agreement within the error bars, defined by the error of the nuclear position in the model and the scatter of the off-axis points in the two extensions with respect to the zero-offset axis.
\begin{figure*}[t]
\begin{center}
\includegraphics[scale=0.5]{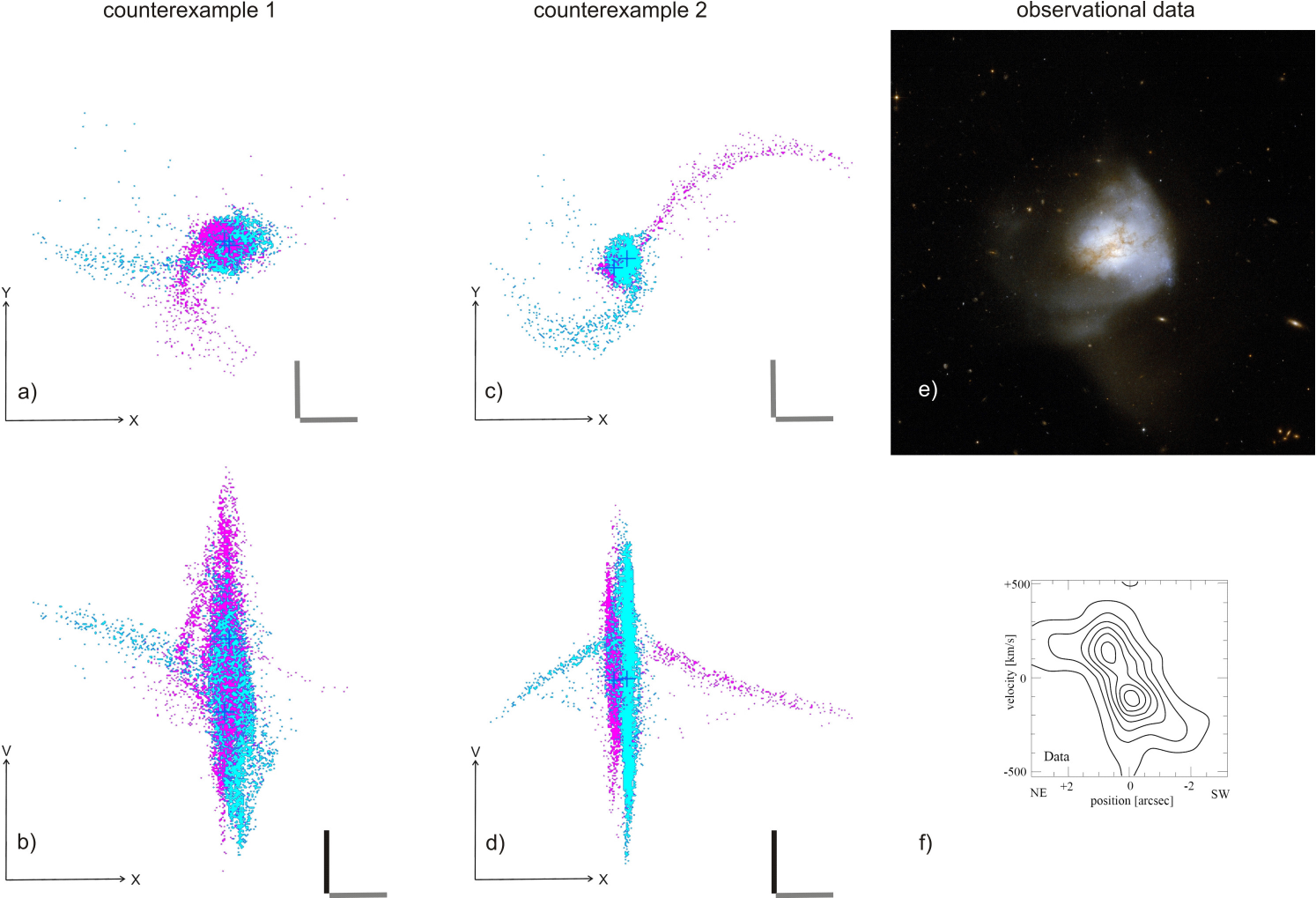}
\caption{\footnotesize Counterexamples to the best model fit. Identikit simulation with the same parameter set as in the best model
fit, but with a mass ratio \textit{$\mu$} of 1:1 (\textit{left}) and a viewing angle \textit{$\Theta_{\rm X}$} of 0\degr\
(\textit{middle}). The x-y plane is shown at the top (a) and c)), the position-velocity diagram (x-v) at the bottom (b)and d)). The
HST composite image and the CO position-velocity diagram \citep[from][]{eck01} used to fit the Identikit simulations are shown in e) 
and f) for comparison. Note that the HST image is rotated by 180\degr\ to match the Identikit internal axis orientation. The grey bar represents a size of roughly 10~kpc and the black bar represents a velocity scale of $\sim$\,500~km\,s$^{\rm -1}$.}
\label{fig:anti_models}
\end{center}
\end{figure*}

\subsection{Discussion} \label{subsection:identikit_discussion}

Since it is known from observations at mm wavelengths that the mass ratio of the two nuclei \object{Arp~220-West} and
\object{Arp~220-East} is about 1:1 \citep[see e.g.,][]{sco97,sak99,dow07}, it is curious that the simulation results presented
here give a value for the mass ratio $\mu$ of the parent galaxies of 1:2. One possible explanation is that the model, is not unique,
i.e. the set of parameters fitting the HST and CO data. 
Nonetheless, the best model obtained here produces a 'fair' fit to the HST
image and the CO data. 
Another point that speaks in the favor, of the model presented here, is the 
determined merger age that is in
very good agreement with the literature value from \citet{mun01}.\\
\indent
However, the situation may be more complicated since (and this can be regarded as a short coming
of our fitting procedure) 
the data that were used to fit the model to, were taken at two different wavelengths
and are therefore basically tracing matter under different physical conditions. 
Only in the central arcseconds the gas has an
impact on the overall gravitational potential, comparable to the stars impact on the potential. Further out the stellar potential
clearly is dominating. Therefore the gas follows the potential of the stars and consequently the stars themselves. Since it was
easier to guide the fits from the simulations by the tidal tails, seen in the HST image, more weight is put on the outer regions
of the galaxy. Hence it is justifiable to use the HST data as well as the CO observations to fit the model results to. Outside the
central region of \object{Arp~220} gas is assumed to be found in clumps with sizes of about 1~pc. Therefore the gas clumps can be
treated, to first order, as stars. Both, stars and gas clumps can then be handled, to first order, as collisionless particles.\\
\indent
To show that the model based on the dynamical multi-particle simulation presented here is indeed one possible qualitative description 
of the system, two possible counterexamples are shown for comparison in Fig.\,\ref{fig:anti_models}. Since a mass ratio
of 1:1 between the nuclei is commonly assumed, one counterexample representing this mass ratio is shown in
Fig.\,\ref{fig:anti_models} (\textit{left}). It was modeled with the same set of parameters as for the best model fit, but with the
different mass ratio $\mu$. In this figure it is very clear that the model does not fit either the HST image nor the CO p-v diagram
(Fig.\,\ref{fig:anti_models}). The same applies to the second counterexample (Fig.\,\ref{fig:anti_models}, \textit{in the middle}).
Here the same set of parameters as for the best model fit was used again, only this time with a viewing angle $\Theta$$_{\rm X}$ of
0\degr, which corresponds to an angular deviation from the best fitting model of 10$\sigma$. The resulting model very clearly does
not fit the underlying observational data.

\section{Conclusions}

In this paper on \object{Arp~220} the results on the study of the interferometric CO data and the Identikit simulations of the
merger are discussed. The eastern nucleus of \object{Arp~220} is studied and the object as a whole is studied for more extended
structure in CO gas emission. The analysis of the CO lower resolution data shows that the velocity dispersion component centered between CO-SE and CO-NE is a very promising candidate to be the 'real', but deeply embedded/highly dust obscured, nucleus of \object{Arp~220-East}. Indications for emission $\sim$\,10\arcsec\ towards the south, as well as to the north and to the west of the two nuclei were found in the low resolution CO\,(1$-$0) maps. Furthermore simulations of the merger in \object{Arp~220} were performed with the Identikit modeling tool. The model parameters describing the galaxy merger best give a mass ratio of 1:2 and result in a merger of $\sim$\,6$\times$10$^{\rm 8}$~yrs, which is in good agreement with values from the literature, such as \citet{mun01}. Apparently \object{Arp~220} can be reckoned a very interesting case of merger galaxy, where the object can be placed between a minor merger (from the CO observations) in one extreme and an equal (1:1) merger (from the simulations, CO data and published models in the literature) in the other.

\acknowledgments

The Dark Cosmology Centre is funded by the Danish National Research Foundation. MG-M is supported by the German federal department for education and research (BMBF) under the project numbers: 50OS0502 \& 50OS0801. Part of this work was supported by the German Deutsche Forschungsgemeinschaft, DFG, via grant SFB956. We thank the anonymous referee for a very careful and constructive report.



{\it Facilities:} \facility{IRAM:Interferometer}, \facility{HST (WFPC2)}.

\clearpage

\end{document}